\documentclass[twocolumn,showpacs,amsmath,prl]{revtex4}
\usepackage{graphicx}
\usepackage{dcolumn}
\usepackage{bm}

\begin{document}

\title{Room temperature spin polarized magnetic semiconductor}

\author{Soack Dae Yoon}
\altaffiliation[Corresponding author e-mail: syoon@ece.neu.edu ]{}
\author{Carmine Vittoria}
\email{vittoria@lepton.neu.edu}
\author{Vincent G. Harris}
\email{harris@ece.neu.edu}
\affiliation{Center for Microwave Magnetic Materials 
and Integrated Circuits, Department of Electrical and 
Computer Engineering, Northeastern University, 
Boston, MA. 02115 USA}

\author{Alan Widom}
\email{widom@neu.edu}
\affiliation{Department of Physics, Northeastern University, 
Boston, MA. 02115 USA}

\begin{abstract}
Alternating layers of granular Iron (Fe) and Titanium dioxide (TiO$_{2-\delta}$) 
were deposited on (100) Lanthanum aluminate (LaAlO$_3$) substrates in low oxygen 
chamber pressure using a controlled pulsed laser ablation deposition technique. 
The total thickness of the film was about 200 nm. The films show ferromagnetic 
behavior for temperatures ranging from 4 to $400 ^oK$. The layered film 
structure was characterized as p-type magnetic semiconductor at $300 ^oK$ with 
a carrier density of the order of $10^{20} /cm^3$. The undoped pure TiO$_{2-\delta}$ 
film was characterized as an n-type magnetic semiconductor. The hole carriers were 
excited at the interface between the granular Fe and TiO$_{2-\delta}$ layers similar 
to holes excited in the metal/n-type semiconductor interface commonly observed 
in Metal-Oxide-Semiconductor (MOS) devices. The holes at the interface were polarized 
in an applied magnetic field raising the possibility that these granular MOS structures 
can be utilized for practical spintronic device applications.
\end{abstract}


\maketitle

\section{introduction}

The search for semiconductors exhibiting magnetism at room temperature has been long and 
unyielding. However, recently much progress has been made toward this 
goal\cite{Ohno:1998,Ueda:2001,Masumoto:2001,Pearton:2003,Chambers:2003,Zutic:2004,Coey:2005,Garcia:2005}. 
By doping a host semiconductor material with transition metal ferromagnetic atoms, dilute 
ferromagnetic semiconductors have been produced with Curie temperatures ($T_c$) as high as 
$160 ^oK$\cite{Ohno:1998}. Hall effect measurements below $T_c$ showed evidence for carriers 
being spin polarized raising hopes for spintronics applications. Specifically, metallic 
Manganese (Mn) was doped into Gallium arsenide (GaAs) whereby approximately $100 \%$ of 
the carriers were spin polarized\cite{Ohno:1998}.

We have reported the magnetic and transport properties of magnetic semiconductor films of 
TiO$_{2-\delta}$, where $\delta$ indicates the degree of oxygen deficiency or defects 
in the film\cite{Yoon1:2006,Yoon:2007}. The Curie temperature, $T_c\approx 880 ^oK$, 
was well above room temperature with a saturation magnetization of $M_s\approx 32$ Gauss. 
Titanium dioxide, TiO$_{2}$, is a well known wideband gap oxide semiconductor, belonging 
to the group IV-VI semiconductors, described in terms of an ionic model of Ti$^{4+}$ and O$^{2-}$\cite{Earle:1942,Breckenridge:1953,Daude:1977,Pascual:1978,Tang:1993}. Its intriguing 
dielectric properties allow its use as gate insulator materials in the Field-Effect-Transistor 
(FET)\cite{Cambell:1999}. Also, TiO$_{2}$ is characterized to be an n-type semiconductor 
with an energy gap varying in the range $3 {\rm \ volt}<\Delta /e< 9 {\rm \ volt}$ depending 
on sample preparation\cite{Earle:1942,Breckenridge:1953,Daude:1977,Pascual:1978,Tang:1993}. 
Films of TiO$_{2-\delta}$  on substrates of (100) Lanthanum aluminate (i.e. LaAlO$_{3}$) 
were deposited by a pulsed Laser-Ablation-Deposition (LAD) technique at various 
oxygen chamber pressures ranging from 0.3 to 400 mtorr. The origin of the presence of 
Ti$^{2+}$ and Ti$^{3+}$ (as well as Ti$^{4+}$) ions was postulated as a result of 
the low oxygen chamber pressure during the films growth\cite{Yoon1:2006}. Oxygen defects 
gave rise to valence states of Ti$^{2+}$ and Ti$^{3+}$ (in background of Ti$^{4+}$) 
whereby double exchange between these sites dominated. The same carriers involved 
in double exchange also gave rise to impurity donor levels accounting for the transport properties 
of the film. The dilute number of carriers were spin polarized, yet the magnetic moment 
was still rather small\cite{Yoon:2007}.For example, normal Hall resistivity was measured 
to be much bigger than any anomalous Hall resistivity\cite{Yoon:2007}. In order to increase 
the anomalous contribution to the Hall effect and thereby increase the number 
of spin polarized carriers, we have fabricated a layered structure of 
TiO$_{2-\delta}$/metallic Iron (Fe). The intent was to introduce a substantial magnetization component 
internally to semiconductor TiO$_{2-\delta}$.

The basic difference between our layered composite and the previously reported magnetic semiconductors 
by others is that in our films the Fe layers co-exist in a metallic state, whereas magnetic 
semiconductors prepared by others the transition metal co-exists as metal oxides and often oxide clusters\cite{Pearton:2003,Chambers:2003,Zutic:2004,Coey:2005,Shinde:2003,Wang:2003}. This major 
difference is important in terms of the magnetic and transport properties of the magnetic semiconductor 
presently produced by us and that of others\cite{Pearton:2003,Chambers:2003,Zutic:2004,Coey:2005,Shinde:2003,Wang:2003}. 
For example, metallic Fe contains a significant higher moment and Curie temperature than any other transition 
metal oxides. Also, the presence of metallic Fe allows for the creation of a reservoir of conduction electrons 
in the conduction band and, therefore, holes in TiO$_{2-\delta}$ layers. As electrons from the conduction band 
of TiO$_{2-\delta}$ are thermally \textquotedblleft {\it dumped} \textquotedblright into the metallic Fe 
conduction band, holes are created in TiO$_{2-\delta}$ much like in junctions of metal/semiconductor interfaces 
or in Metal Oxides Semiconductor (MOS) devices. Conduction of holes occurs in TiO$_{2-\delta}$ layers. This 
mechanism gives rise to lower resistivity at high temperature in contrast to pure TiO$_{2-\delta}$ reported 
earlier\cite{Yoon1:2006} where the carriers were only electrons. No such mechanism is possible in magnetic 
semiconductors doped with transition metal oxides. We report here that in our layered semiconductor of 
Fe/TiO$_{2-\delta}$ the magnetization, ${M_s}\approx 250 $, at room temperature, $T_c$ above $800 ^oK$, 
nearly $100 \%$ of the carriers are spin polarized and the room temperature resistivity is lowered by as 
factor of $\sim 10^{-3}$ relative to films of the undoped TiO$_{2-\delta}$  semiconductors previously 
produced\cite{Yoon1:2006}. 

\begin{figure}[bp]
\centering
\includegraphics[width=0.395\textwidth]{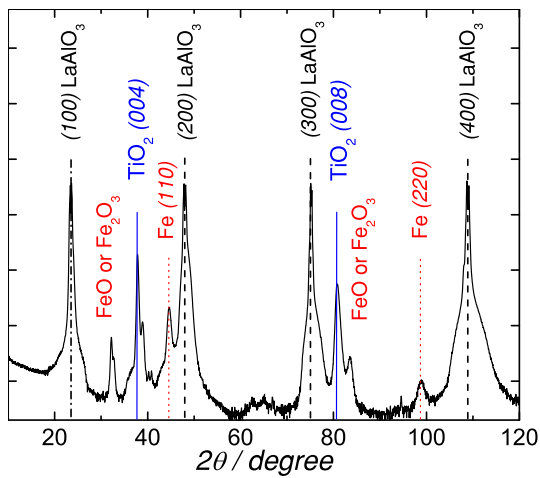}
\caption{X-ray diffraction spectrum for a representative 
layered film of granular Fe in TiO$_{2-\delta}$.}
\label{XRD} 
\end{figure}

\section{Experimental Process}
Thin films consisting of alternating layers of granular iron (Fe) and Titanium dioxide (TiO2- ) were deposited 
by a pulsed laser ablation deposition (LAD) technique from binary targets of TiO2 and metallic Iron (Fe) on (100) 
Lanthanum aluminate (LaAlO3) substrates. We refer to this technique as the Alternating Targets LAD 
(or ALT-LAD)\cite{Zuo:2005,Yoon2:2006}. Alternate layers of granular Fe and TiO$_{2-\delta}$ were deposited 
sequentially to produce films whose crystal structure was aimed to be similar to that of pure TiO$_2$. Targets 
of TiO$_2$ and Fe were mounted on a target rotator driven by a servomotor and synchronized with the trigger of 
the pulsed excimer laser, $\lambda$ =248 nm. In each deposition cycle, the ratio of laser pulses incident upon 
the TiO$_2$ target to those upon the Fe target was 6:1. The substrate temperature, laser energy density, and 
pulse repetition rate were maintained at $700 ^oC$, $\approx 8.9 {\rm J /cm^2}$, and 1 Hz, respectively. The 
deposition was carried out in a pure oxygen background of around 10  torr in order to induce defects in the TiO$_2$ 
host. There were a total of 4200 laser pulses (3600 pulses on TiO2 target and 600 pulses on Fe target) for each film 
resulting in a thickness of approximately 200 nm as measured by a Dek-Tek 3 step profilometer. Crystal and surface 
structure measurements were performed by x-ray diffractometer (XRD) and Scanning-Electron-Microscopy (SEM). 
In order to measure the ratio of Ti to Fe, method of Energy Dispersion X-Ray Spectroscopy (EDXS) within the SEM 
column was employed. Magnetic hysteresis loops and dc-electrical and magneto-resistivity of the resultting films 
were measured using a Quantum Design Physical Property Measurement System (PPMS) in the temperature range between 
$4 ^oK$ and $400 ^oK$.  

\begin{figure}[bp]
\centering
\includegraphics[width=0.43\textwidth]{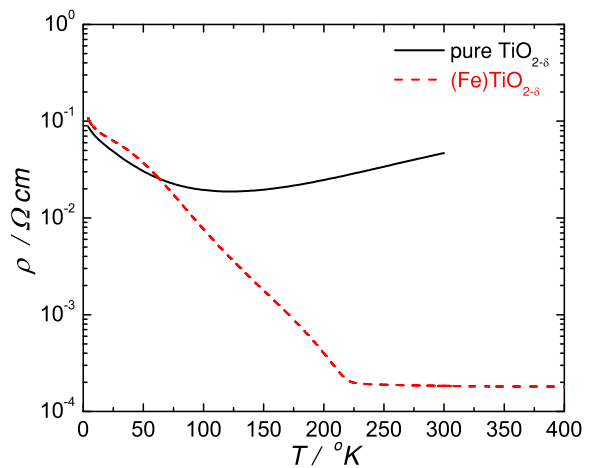}
\caption{Resistivity, $\rho$ as a function of temperature, $T$, 
for the layered granular Fe in TiO$_{2-\delta}$ film (dashed line) 
and the pure TiO$_{2-\delta}$ (solid line) from the 
reference\cite{Yoon1:2006}}
\label{Resistivity} 
\end{figure}

\section{Results and Discussions}
Crystallographic properties of the AT-LAD films were measured using x-ray diffractometry (XRD). 
Results indicate multiple phases of TiO$_2$, metallic Fe, and iron oxide (Fe$_2$O$_3$). FIG.\ref{XRD} shows 
a representative XRD spectrum of the film. The diffraction peaks at $2\theta = 37.76^o$ and $80.714^o$ were 
indexed to the $\it(004)$ and $\it(008)$ planes, respectively, of the Anatase TiO$_{2-\delta}$TiO$_{2-\delta}$. 
The metallic Fe phase was identified by the diffraction peaks appearing at 2$\theta = 44.643^o$ and $98.929^o$
which are indexed to the $\it(110)$ and $\it(220)$ planes\cite{Swanson:1955}, respectively. Fe$_2$O$_3$ peaks 
were also observed in the spectrum implying that the embedded granular Fe was possibly partially oxidized.

For steady currents and with the magnetic intensity ${\bf H}$ aligned normal to the film plane, the resistance 
matrix ${\sf R}$ in the plane of the film may be written as\cite{Landau:1982} 
\begin{equation}
{\sf R}=
\begin{pmatrix}
R_{xx} & R_{xy} \\ 
R_{yx} & R_{yy}
\end{pmatrix}
=\frac{1}{t}
\begin{pmatrix}
\rho & -\rho_H \\ 
\rho_H & \rho
\end{pmatrix}
\label{transport1}
\end{equation}
wherein $\rho $ and $\rho_H$ represent, respectively, the normal-resistivity and the Hall resistivity. 
A $\rho$ as a function of temperature for the layered film was measured in an applied field of $H = 0 Oe$ and shown 
in FIG.\ref{Resistivity}. We note that the value of $\rho(T)$ was quite small at high temperature. The $\rho(T)$
behavior for pure TiO$_{2-\delta}$ film 10 (solid line) is also shown in FIG.\ref{Resistivity} exhibiting a typical 
metal-insulator transition in the temperature range of $4^oK$ and $300^oK$. In contrast to the $\rho$ for pure 
TiO$_{2-\delta}$, $\rho$ for the layered film is a factor of 1000 lower and is constant for temperatures between 
$225^oK$ and $400 ^oK$. The $\rho$ value at $300 ^oK$ was measured to be 183 $\mu\Omega$$\it cm$, and is about 
a factor of 20 larger than the resistivity of pure Fe\cite{Weast:1982}.

\begin{figure}[tp]
\centering
\includegraphics[width=0.44\textwidth]{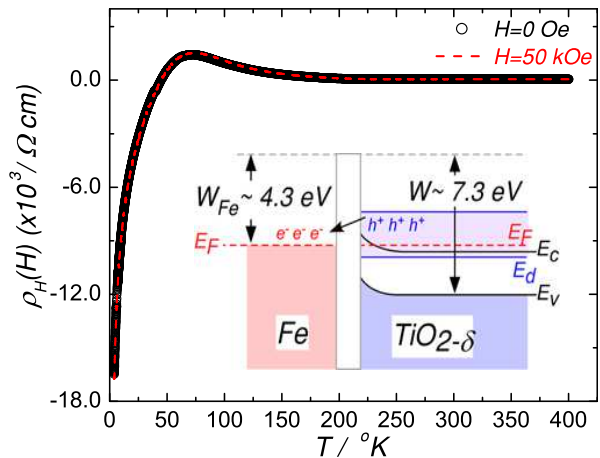}
\caption{Hall resistivity, $\rho_H$, versus temperature for $H = 0$ (circle symbol) 
and $H = 50\ kOe$ (dashed line) in the layered granular Fe in TiO$_{2-\delta}$ film. 
Inset figure is an energy model of a metal oxide semiconductor}
\label{HallResistivity} 
\end{figure}

We have modeled the mechanism for transport in this layered film in a sketch shown in inset of 
FIG.\ref{HallResistivity}\cite{Streetman:1990,Knotek:1978,Zhang:1991}. A common chemical potential energy or 
Fermi energy in the layered film implies that the conduction band of TiO$_{2-\delta}$ layers is degenerate 
with the metallic Fe conduction band. Electrons hopping between Ti$^{2+}$ and Ti$^{3+}$ sites would find 
a conduit into the metallic conduction band whereby holes are created in the TiO$_{2-\delta}$ layers and therefore 
a small potential barrier at the interface. This is illustrated schematically in FIG.\ref{HallResistivity}. Since, 
the Fe metallic particles are isolated or disconnected, conduction in the TiO$_{2-\delta}$ layers is via 
hole conduction. At very low temperatures, where thermal kinetic energy is not sufficient to raise conduction electrons 
into the conduction band of the Fe particles, conduction is mostly due to electron carriers in TiO$_{2-\delta}$ layers. 
In FIG.\ref{HallResistivity}, Hall resistivity is plotted as a function of temperature. Indeed, at high temperature 
carriers are holes and electrons for $T < 50 ^oK$. Furthermore, the number of carriers relative to pure 
TiO$_{2-\delta}$ has increased by factor 1000 and the mass of holes is approximately about 10 times larger than 
the electron mass. Effectively, the Hall resistivity in the composite layered structure is decreased by a factor of 
1000 relative to pure TiO$_{2-\delta}$. The experimental data at $300 ^oK$ shows $\rho_{H} = 34.00 m\Omega\ {\rm cm}$ 
for pure TiO$_{2-\delta}$ and $\rho_{H} = 53.81 \mu\Omega\ {\rm cm}$ for layered granular Fe TiO$_{2-\delta}$ 
(or approximately a factor of 630 smaller). 

\begin{figure}[tp]
\centering
\includegraphics[width=0.43\textwidth]{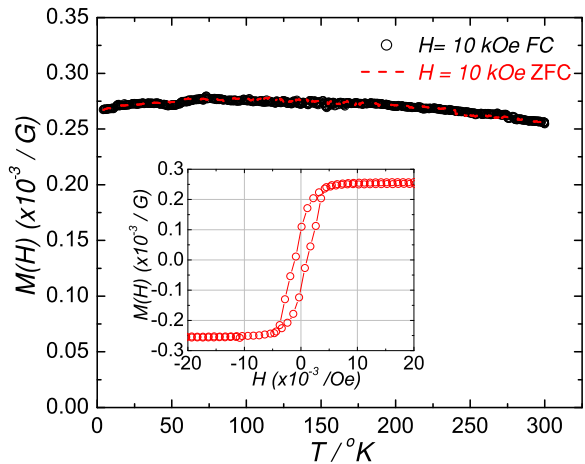}
\caption{Static magnetization, $M(H,T)$, versus temperature with $H =10\ kOe$
normal to the film plane. Inset figure is a magnetic hysteresis loop measured at 
$T =305 ^oK$.}
\label{Magnetization} 
\end{figure}

The spontaneous magnetization was measured to be nearly constant as a function of temperature as shown 
in FIG.\ref{Magnetization}. The measurement was performed with an external dc-magnetic field of $10 kOe$ applied normal 
to the film plane (out-of-plane measurement). The out-of-plane magnetic hysteresis loop behavior at $305^oK$ is shown 
in the inset of FIG.\ref{Magnetization}. The film can be fully saturated with a field of $10 kOe$, since the coercive field 
was measured to be $2 kOe$. Field cooled (FC) and zero-field-cooled (ZFC) $M(H,T)$ data in FIG.\ref{Magnetization} shows 
no difference which implies that no spin glass effects are present in the films. The magnetization ($M_s$) of pure 
TiO$_{2-\delta}$ was measured to be about T$M_s \approx 32$ Gauss at $300^oK$ which is considerably smaller than that 
measured for the Fe/TiO$_{2-\delta}$ films. Clearly, the measured magnetization of the layered films must be due to 
the presence of metallic Fe.

\begin{figure}[tp]
\centering
\includegraphics[width=0.44\textwidth]{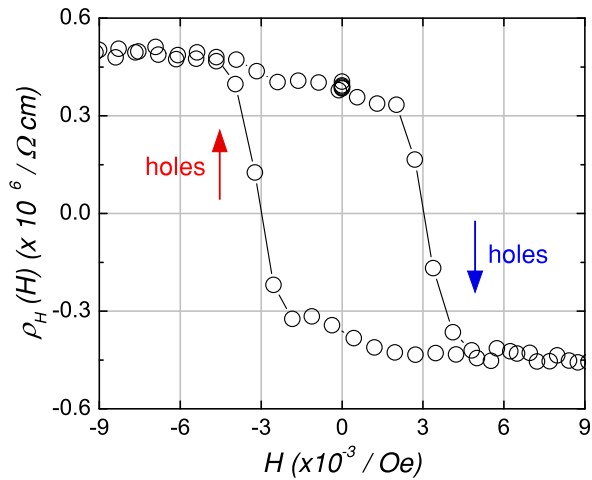}
\caption{Hall resistivity, $\rho_H$, versus $H$ at $300^oK$. 
Carrier polarization density is of the order of $10^{20}/cm^{3}$ with 
a Hall resistivity value of 0.8 $\mu\Omega\ {\rm cm}$.}
\label{HystereticHall} 
\end{figure}

Finally, we have measured an anomalous Hall effect in magnetic field sweeps between -90 and $90 kOe$ as shown 
in FIG.\ref{HystereticHall}. The film exhibited strong polarization and hysteresis behavior as a function of applied field 
at $300 ^oK$. Since, the anomalous Hall effects may be observed in the presence of a spontaneous 
magnetization\cite{Chien:1980,Karplus:1954} we infer that about $10^{20}/cm^{3}$ hole carriers are spin polarized. 
This polarized carrier density value is a factor of 1.2 larger than the carrier density estimated from normal Hall 
measurement, where carrier density was estimated with only applied $\textbf H$, at $300 ^oK$. This indicates that 
nearly all of the carriers were spin polarized by the spontaneous magnetization. FIG.\ref{HystereticHall} indicates 
that the carrier polarization is not affected by external field up to $3 kOe$, which can be an advantage for 
memory device applications\cite{Prinz:1998,Wolf:2001}.

\section{Conclusions}
Magnetic and magneto-transport data for layered films of metallic iron (Fe) and oxygen defected titanium oxide 
(TiO$_{2-\delta}$) are reported in this letter. The essence of this paper showed that conduction carriers of 
the films were strongly coupled to residual magnetic moments of metallic iron (Fe) layers in the layered structure. 
The dramatic reduction of normal resistivity ($\rho(T)$) of the films is a consequence of two factors: (1) oxygen defects 
in the TiO$_{2-\delta}$ layers induced electron hopping; (2) electrons from the TiO$_{2-\delta}$ were \textquotedblleft 
{\it dumped} \textquotedblright into the conduction band of Fe layers to create holes in TiO$_{2-\delta}$ similar to 
a Metal Oxide Semiconductor (MOS) structure. As a result the number of carriers increased, and at room temperature, 
the majority carriers were holes with a density of $5 \times 10^{20} /cm^{3}$ as measured by normal Hall measurements. 
The holes in TiO$_{2-\delta}$ were polarized due to the presence of ferromagnetic metallic Fe, where the spin polarized 
density was measured to be $6 \times 10^{20} /cm^{3}$. Therefore, spintronics and spin dependent memory applications 
can be based upon the results presented here.

\end{document}